\documentclass{aa}
\usepackage{natbib}
\bibpunct{(}{)}{;}{a}{}{,}

\usepackage{epsfig}
\usepackage{graphics}
\usepackage{float}
\usepackage{amsmath}
\usepackage{multirow}
\usepackage{longtable}
\usepackage{rotate}
\usepackage{array}
\usepackage{subfigure}
\def\sidehead#1{\noalign{\vskip 1.5ex}\multicolumn{4}{@{}l}{\em #1}\\
                \noalign{\vskip .5ex}}

\def\phn{\phantom{0}}  
\def\phs{\phantom{$-$}}    

\def\tablecomments#1{\par\smallskip\noindent Notes. #1}
\def\plotone#1{\centerline{\psfig{figure=#1,width=\hsize,clip=}}}
\def\kms{\ifmmode{\rm km\,s^{-1}}\else\hbox{$\rm km\,s^{-1}$}\fi}

\newcommand{\Teff}{$\mathrm{T}_{\mathrm{eff}}$}

\let\arcdeg\degr

\setlongtables

\begin{document}

\title{A puzzling periodicity in the pulsating DA white dwarf G\,117-B15A
      \thanks{Based in part on data obtained at
       the W.M. Keck Observatory, which is operated as
       a scientific partnership among the California Institute of Technology,
       the University of California and the National Aeronautics and Space
       Administration. The Observatory was made possible by the generous
       financial support of the W.M. Keck Foundation.}}

 \author{R. Kotak\inst{1,2}
          \and{M. H. van Kerkwijk}\inst{3,4}
           \and{J. C. Clemens}\inst{5}\thanks{Alfred P. Sloan Research Fellow}}
   \offprints{R. Kotak}

   \institute{ Lund Observatory,
               Box 43, SE-22100 Lund, Sweden
           \and
               Astrophysics Group, Imperial College London, Blackett Laboratory, 
               Prince Consort Road, \\ London, SW7 2BZ, U.K.  \,\,\,
               \email{rubina@ic.ac.uk}
           \and
              Astronomical Institute, Utrecht University,
              P. O. Box 80000, 3508~TA Utrecht, The Netherlands 
           \and
              Department of Astronomy and Astrophysics, University of Toronto,
              60 St George Street, Toronto, \\ Ontario M5S 3H8, Canada \,\,\,
              \email{mhvk@astro.utoronto.ca}
           \and
              Department of Physics and Astronomy, University of
              North Carolina, Chapel Hill, NC 27599-3255, USA\\
              \email{clemens@physics.unc.edu}
             }
   \date{Received 30 June 2003; Accepted 22 September 2003}

   \abstract{We present time-resolved optical spectrophotometry of the pulsating
            hydrogen atmosphere (DA) white dwarf G 117-B15A. We find three
            periodicities in the pulsation spectrum (215\,s, 272\,s, and 304\,s)
            all of which have been found in earlier studies. By comparing the
            fractional wavelength dependence of the pulsation amplitudes (chromatic
            amplitudes) with models, we confirm a previous report that the strongest
            mode, at 215\,s, has $\ell=1$. The chromatic amplitude for the 272\,s 
            mode is very puzzling, showing an increase in fractional amplitude with 
            wavelength that cannot be reproduced by the models for any $\ell$ at 
            optical wavelengths. Based on archival {\em HST\/} data, we show that 
            while the behaviour of the 215\,s mode at ultra-violet wavelengths is 
            as expected from models, the weird behaviour of the 272\,s periodicity 
            is not restricted to optical wavelengths in that it fails to show the 
            expected increase in fractional amplitude towards shorter wavelengths. 
            We discuss possible causes for the discrepancies found for the 
            272\,s variation, but find that all are lacking, and conclude that the 
            nature of this periodicity remains unclear.
   }
  \authorrunning{Kotak}
   \titlerunning{A puzzling periodicity in G 117-B15A}
   \maketitle

\section{Introduction}

The hydrogen-atmosphere pulsating white dwarfs (DAVs or ZZ Cetis), 
occupy a narrow (1000\,K wide) instability strip at $\sim$\,11.5\,kK, 
and exhibit multi-periodic flux variations of several hundreds of 
seconds that are primarily due to changes in the effective temperature
\citep{rkn:82}.
The DAVs can be roughly divided into the hotter objects that have
relatively simple and stable pulsational spectra consisting of a few
short period, low amplitude modes, and the cooler ones that have more 
modes in total, with generally longer periods and larger amplitudes, 
and show moderate to severe amplitude variability on several different 
timescales, with some modes even disappearing from one season to the 
next \citep[e.g.][]{koester:02}. 

In order to infer the interior properties of the white dwarf, the modes 
in any given pulsator must be individually identified i.e. the spherical 
degree ($\ell$; number of nodal lines on the surface), the azimuthal order
($m$), and the radial order ($n$; number of nodes from centre to surface) 
have to be known with some confidence. While $\ell$ and $m$ can be determined 
observationally, $n$ has to be deduced by detailed comparison with pulsation 
models. The spherical degree can be inferred in a number of ways, none of 
which are completely reliable on their own. Most studies to date have relied 
on the rotationally-induced splitting of modes and/or the comparison of 
observed distribution of mode periods with predicted ones. Assuming spherical 
symmetry, frequencies of modes having the same $n$ and $\ell$ are degenerate. 
Slow rotation lifts this degeneracy resulting in $2\ell+1$ split components. 
The observation of all split components is an indicator of the $\ell$ value 
of a mode. The other method relies on the expectation that in the absence 
of compositional boundaries and for large $n$, modes of consecutive radial 
order (and same $\ell$) are equally spaced in period.

\citet{robetc:95} developed another method which relies on the 
increased importance of limb-darkening at short wavelengths and the 
effect this has on pulsation amplitudes as a function of wavelength. 
The advantage of this method is that the resulting variation of mode 
amplitudes with wavelength is a function of $\ell$ (but not $m$), 
thereby permitting mode-identification in the absence of a large 
number of modes or rotationally-split modes. We will focus on this 
latter method here.

G 117-B15A epitomises the hotter DAVs, and has been extensively observed 
since the discovery of its variability almost three decades ago 
\citep{richul:74,mcgrob:76}. Its pulsation spectrum is relatively simple, 
with the dominant mode occurring at 215\,s. This mode has been shown to 
be remarkably stable in amplitude \citep{kepler:00a}. Two other modes at
272 and 304\,s also appear in all data sets, although they are weaker
and show variations in amplitude. Other possible real modes have also 
been reported by \citet{kepler:95}. However, the three modes mentioned 
above appear in all data sets for this object. A Whole Earth Telescope 
\citep[WET,][]{nath:90} run failed to reveal both a coherent set of 
splittings for any mode \citep{kepler:95}, and a series of modes that 
could be identified with a particular $\ell$ and consecutive $n$.
A clear $\ell$ identification was therefore not possible. 

On the other hand, ultra-violet light curves and quantitative use of model 
atmospheres allowed \citet{robetc:95} to assign $\ell=1$ to the 215\,s mode in 
G 117-B15A. Using this $\ell=1$ identification of the 215\,s mode, and {\em assuming\/} 
an $\ell=1$ identifcation for the 272\,s and 304\,s modes, \citet{bradley:98} 
derived a range of parameters, from thickness of the superficial hydrogen layer 
to the core \ion{C}/\ion{O}\ ratio, from fits to pulsation models. 

The primary motivation for the work described in this paper stemmed from a 
need for independent confirmation of the $\ell$ identification of the 215\,s 
mode, and the need for constraints on the $\ell$ values of the other modes. 
As we describe below, our data however, have revealed a number of surprises.

\begin{figure}
\plotone{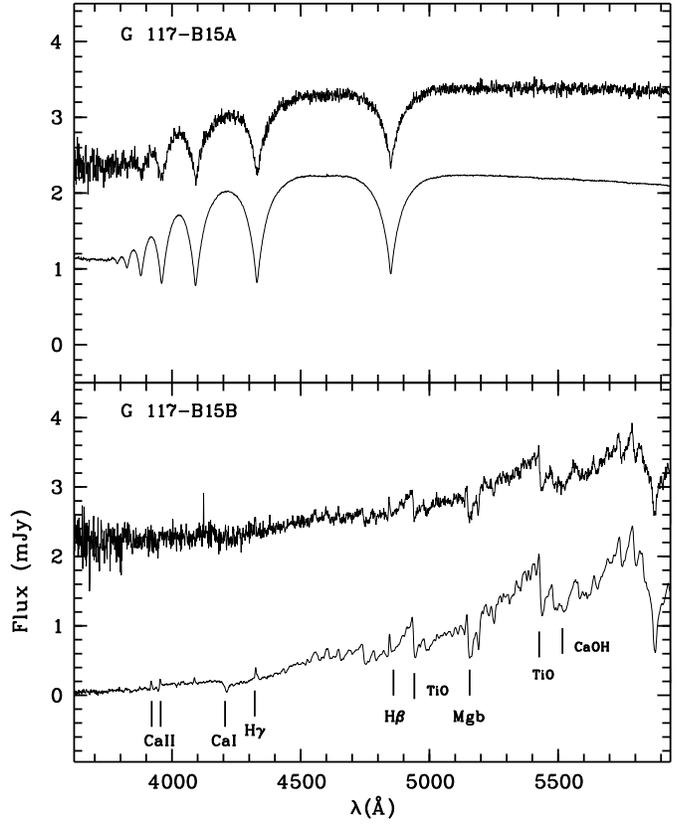,angle=-90}
\caption{Sample and average spectra of G 117-B15A (top panel) and its common 
proper motion companion G 117-B15B (bottom panel). The sample spectrum of 
G 117-B15A is offset from the mean spectrum by +1.7\,mJy while that of 
G 117-B15B is offset by +2.2\,mJy. The average spectra of G 117-B15A 
and G 117-B15B  have been scaled such that the fluxes at $\sim$\,5500\,{\AA} 
correspond to the published V band magnitudes (V=+15.54 and +16.06 respectively).}
\label{fig:avspec}
\end{figure}

\section{Observations}

Time-resolved spectra of G 117-B15A were obtained using the Low
Resolution Imaging Spectrograph \citep{oke:95} mounted at the Cassegrain
focus of the Keck II telescope on 1997 December 11. We acquired a total 
of 235 frames, from 10:47:42 to 14:02:56 U.T. An 8$\farcs$7-wide slit 
was used together with a 600 line\,mm$^{-1}$ grating covering the 
3450-5960\,{\AA} range at 1.25\,\AA\,pixel$^{-1}$. The seeing was 
1\farcs2 and some patchy cirrus was present. The common proper motion 
companion was also accomodated in the slit. The reduction of the data 
included the usual procedures of bias subtraction, flat fielding, optimum
extraction, and wavelength and flux calibration. Flux calibration was
carried out with respect to the flux standard G 191-B2B. The use of a
wide slit (for photometric accuracy) requires additional corrections
to the wavelength scale for differential atmospheric refraction and 
wandering in the slit. These were performed in a manner identical to 
that described in \citet{kotakhs:02}. 

To complement the optical data, we retrieved archival data taken at
ultra-violet wavelengths with the High Speed Photometer on-board the 
{\em Hubble Space Telescope\/} \citep[see][]{robetc:95}. The target 
was observed four times with the F184W filter ($\lambda_{\mathrm{eff}}=
1920$\,{\AA}, FWHM = 360\,{\AA}) and twice with the F145M filter 
($\lambda_{\mathrm{eff}}=1570$\,{\AA}, FWHM = 215\,{\AA}). 

\section{Average Spectra}

Sample and average spectra of G 117-B15A,B are shown in Fig. \ref{fig:avspec}.
As expected, the spectrum of G 117-B15A shows Balmer lines only. 
The spectrum of the common proper motion companion, \object{G 117-B15B}, 
shows strong TiO bands indicating that its spectral type is later than 
about M0. The TiO band head at $\sim$4750\,{\AA} is not particularly 
well-developed due to a blend with the MgH $\lambda 4780$\,{\AA} feature, 
suggesting a spectral type of M3V-M4V. The presence of the CaOH band
and the absence of the \ion{Fe}{i}\ $\lambda$4384\,{\AA} line also suggests 
a spectral type of M3V-M5V. H$\beta$, H$\gamma$, and possibly H$\delta$ are
seen in emission as are the weak \ion{Ca}{ii}\ H and K lines. We
tentatively assign a spectral type of M3Ve.

Attempting to match the flux at $\sim$5500\,{\AA} and the slope of the 
spectrum using the NextGen models described in \citet{hauschildt:99},
we obtain a reasonable match for \Teff $\sim 3.4$\,kK and $\log g \sim 4.5$
with somewhat higher metallicity models, [M/H]=0.5, providing a slightly better 
match than models computed with solar metallicity i.e. [M/H]=0.0.

\section{Periodicities in the light curves}
\label{sec:periodlc}

\begin{figure}[!t]
\plotone{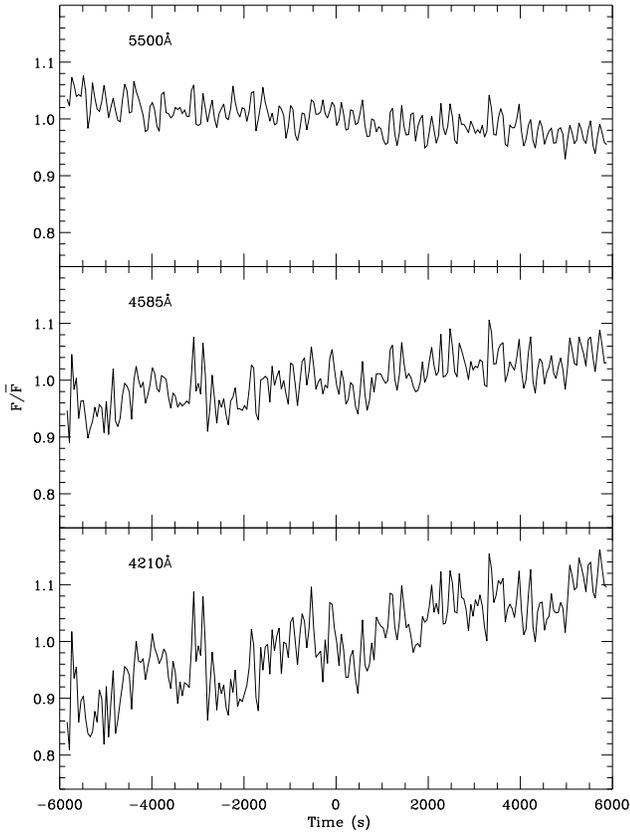}
\caption{Light curves constructed from the optical spectra.
The times shown are relative to the middle of the time series.}
\label{fig:goptlcs}
\end{figure}

\begin{figure}[!t]
\begin{center}
\mbox{
     \includegraphics[width=0.5\textwidth,clip=]{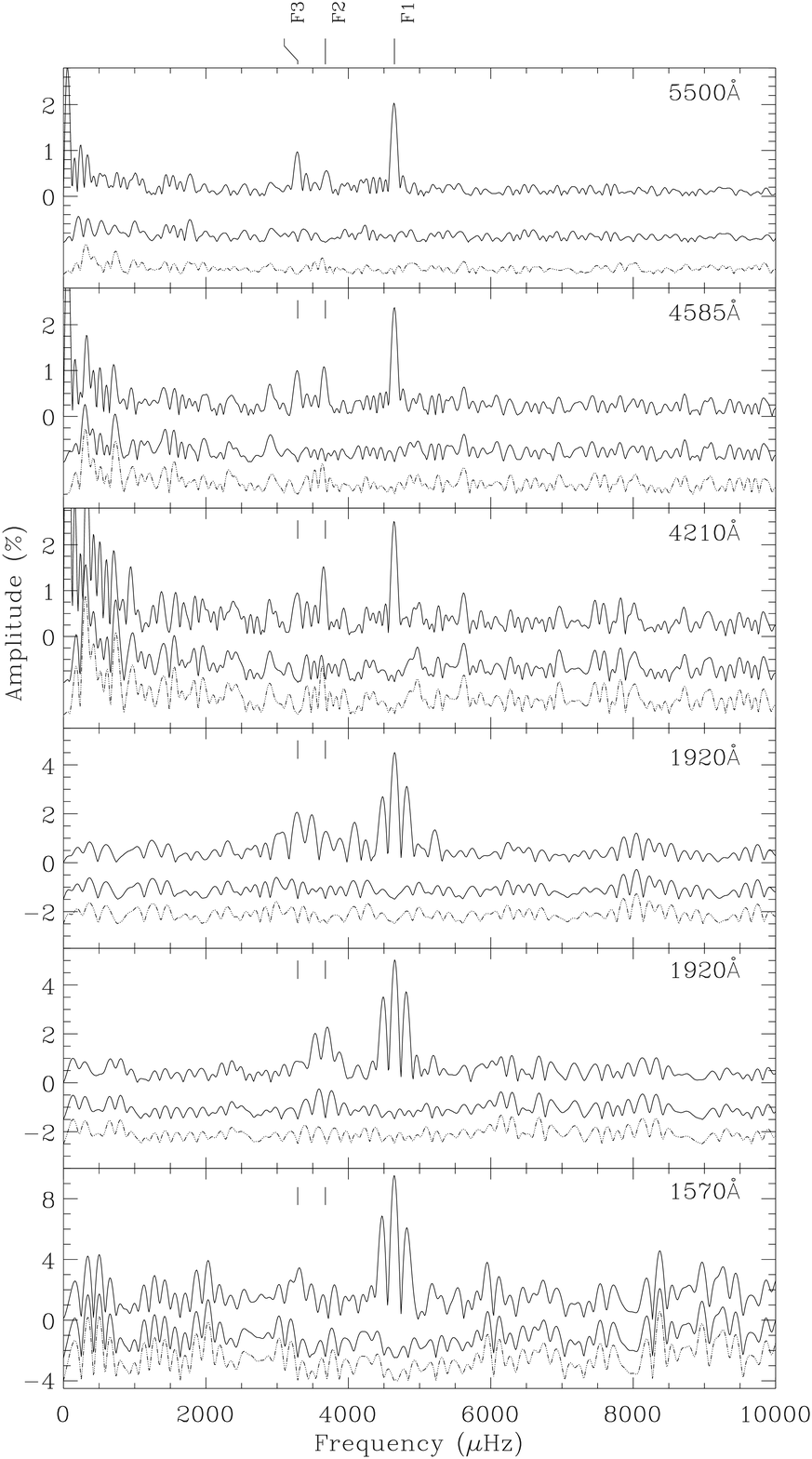} 
     }
\end{center}
\caption{Fourier Transforms (FTs) of the light curves for data taken in
different passbands, with the approximate central wavelengths shown in the
top righthand corner of each panel. The top three panels show the FTs from
our optical dataset while the lower three show the FTs from archival UV data.
The middle curve in each panel shows the residuals after fitting sinusoids
with frequencies listed in Table \ref{gtab}. These are offset by $-$1\% in
the top three panels, by $-$1.5\% in the next two panels, and by $-$2.5\% in 
the lowermost panel. The dash-dot-dot-dashed (lowermost) curve in all panels 
shows the residuals (offset by $-$1.5\% in the top three panels, by $-$2.5\% 
in the next two panels, and by $-$4\% in the lowermost panel) obtained by 
including a periodicity at 270.455\,s instead of one at 271.95\,s (F2). 
Fig. \ref{fig:F2cfperiods} shows an expanded view. Larger residuals result 
in every case. The positions of F2 and F3 are marked in all panels. Note 
the differing scales of the vertical axes of the two sets of triplots.}
\label{fig:glcft}
\end{figure}

\begin{figure}[htb]
\plotone{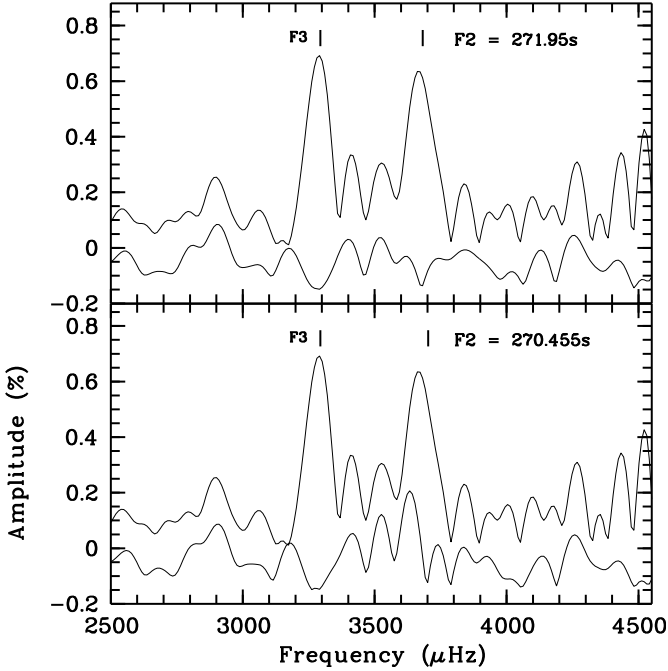}
\caption{Expanded view of the Fourier Transforms of the 5500\,{\AA} light 
 curve around the region containing F2 and F3. The lower curve in both
 panels shows the residuals (offset by $-$0.15\%) as obtained by using the
 periods indicated for F2. The choice of 271.95\,s for F2 provides a clearly
 better fit.}
\label{fig:F2cfperiods}
\end{figure}

\begin{table}[!t]
\caption[]{Amplitudes and phases of the modulations derived from the
   optical and ultra-violet light curves.}
 \begin{centering}
 \label{gtab}
 \begin{tabular}{cccccccc}
\hline 
 Mode & Period  &  Frequency & $A_L$         & $\Phi_L$           \\
      & (s)     &  ($\mu$Hz) & (\%)          &  (\arcdeg)         \\ 
\hline 
\sidehead{Optical (5500\,{\AA})}
   F1 &  215.20 &  4646.9 & 2.01 $\pm$ 0.13  &  $-$153    $\pm$ \phn4 \\
   F2 &  271.95 &  3677.1 & 0.63 $\pm$ 0.13  &  \phs107   $\pm$ 12    \\
  (F2 &  270.46 &  3697.5 & 0.58 $\pm$ 0.13 & \;\phs109   $\pm$ 13)  \\
   F3 &  304.05 &  3288.9 & 0.96 $\pm$ 0.13  &  \phn$-$55 $\pm$ \phn8 \\
\sidehead{Optical (4585\,{\AA})}
   F1 &   215.20 &  4646.9 & 2.37 $\pm$ 0.23 & $-$152     $\pm$ \phn6  \\
   F2 &   271.95 &  3677.1 & 0.99 $\pm$ 0.23 & \phs\phn94    $\pm$ 13  \\
  (F2 &   270.46 &  3697.5 & 0.77 $\pm$ 0.23 &\;\phs\phn96   $\pm$ 17)  \\
   F3 &   304.05 &  3288.9 & 1.07 $\pm$ 0.23 &  \phn$-$50 $\pm$ 12  \\
\sidehead{Optical (4210\,{\AA})}
   F1 &   215.20 &  4646.9 &  2.50 $\pm$ 0.37 &  $-$153      $\pm$ \phn8  \\
   F2 &   271.95 &  3677.1 &  1.22 $\pm$ 0.37 &  \phs\phn97  $\pm$ 17  \\
  (F2 &   270.46 &  3697.5 &  0.83 $\pm$ 0.37 &  \;\phs\phn96  $\pm$ 26)  \\
   F3 &   304.05 &  3288.9 &  1.08 $\pm$ 0.37 &  \phn$-$39   $\pm$ 20  \\ 
\sidehead{UV; v3 (1920\,{\AA})}
   F1 & 215.20 & 4646.9 & 4.56 $\pm$ 0.33 & $-$55 $\pm$ \phn4  \\
   F2 & 271.95 & 3677.1 & 1.61 $\pm$ 0.33 & $-$20 $\pm$ 12    \\
  (F2 & 270.46 & 3697.5 & 1.72 $\pm$ 0.33 & \;\phn$-$16 $\pm$ 11)   \\
   F3 & 304.05 & 3288.9 & 2.01 $\pm$ 0.33 & $-$20 $\pm$ \phn9    \\
\sidehead{UV; v4 (1920\,{\AA})}
   F1 & 215.20 & 4646.9 & 5.04 $\pm$ 0.32 & \phs152 $\pm$  \phn4  \\
   F2 & 271.95 & 3677.1 & 2.00 $\pm$ 0.32 & $-$128  $\pm$  \phn9  \\
  (F2 & 270.46 & 3697.5 & 2.27 $\pm$ 0.32 & \;$-$124  $\pm$  \phn8)  \\
   F3 & 304.05 & 3288.9 & 1.19 $\pm$ 0.33 & $-$157  $\pm$   16  \\
\sidehead{UV; v1 (1570\,{\AA})}
   F1 & 215.20 & 4646.9 & 9.51 $\pm$ 1.34 & $-$128    $\pm$ \phn8  \\
   F2 & 271.95 & 3677.1 & 1.43 $\pm$ 1.34 & \phn$-$66 $\pm$ 54     \\
  (F2 & 270.46 & 3697.5 & 1.25 $\pm$ 1.34 & \;\phn$-$95 $\pm$ 62)   \\
   F3 & 304.05 & 3288.9 & 3.12 $\pm$ 1.34 & $-$153    $\pm$ 24     \\ 
\hline
\end{tabular}
\tablecomments{$A_L$ and $\Phi_L$ are the amplitudes and phases
 obtained by iterative least squares fitting of the light curves.
 The periods have been fixed to those determined from WET data
 \citep{kepler:95} and with a precision higher than that listed
 above: F1 = 215.1968\,s, F2 = 271.95\,s, F3 = 304.052\,s.
 As such, in the ultra-violet, the measurements above represent limits 
 to the amplitudes at a given frequency when the signal is not clearly
 detected (see Fig. \ref{fig:glcft}).
 Also shown in parenthesis, are the values obtained for the 270.46\,s
 periodicity which was assigned a lower probability of being
 a false peak by \citet{kepler:95}.
 Limits for 119.84\,s (8344.7\,$\mu$Hz), a potential real mode
 reported by \citet{kepler:95} are: v1 $4.08\pm1.34$\%
 v3 $0.72\pm0.33$\%, v4 $0.94\pm0.32$\%; optical (5500\,{\AA}): 
 $0.09\pm0.11$\%. The ``v'' numbers merely denote abbreviated run
 numbers.}
\end{centering}
   \end{table}

Optical light curves were constructed by dividing the line-free region
of the continuum between $\sim$\,5300\,-\,5700\,{\AA} (longward of H$\beta$),
$\sim$\,4400\,-\,4800\,{\AA} (between H$\gamma$ and H$\beta$), and
$\sim$\,4000\,-\,4300\,{\AA} (between H$\epsilon$ and H$\gamma$) by a portion
of the spectrum of the companion stretching from $\sim$\,4150\,-\,5470\,{\AA}.
This was done in order to cancel out atmospheric fluctuations.
We preferred the above procedure rather than dividing each chosen region
of G 117-B15A by an exactly equivalent region of G 117-B15B as there is very 
little flux from G 117-B15B shortward of $\sim$\,4000\,{\AA}. This procedure 
does, however, introduce a slight linear trend in the light curves as a result
but this is easily accounted for by including a first-order polynomial in the 
fit of the light curves as described below.

Fig. \ref{fig:glcft} shows the Fourier Transforms of the normalised light 
curves for both the three optical and ultra-violet datasets, and the residuals 
after fitting the sum of three sinusoids of the form $A(\cos 2\pi ft - \phi)$.
Here, $A$ denotes the amplitude, $f$ the frequency, and $\phi$ the phase. 
As mentioned earlier, a low order polynomial was also included in the fit. 
The frequencies were fixed to those reported by \citet{kepler:95} from WET 
data as these are more precise. Given their higher frequency resolution, 
\citet{kepler:95} found several peaks in each of the 215, 270, and 304\,s 
regions. For each of our three peaks, our choice of periods from their
Table 2 was based on the one that resulted in the best fit of our light curve(s). 
For two of the peaks, {\em viz.\/} 215\, and 304\,s, these also corresponded 
to the ones to which they attached the highest significance. We therefore fixed 
these to 215.1968\,s and 304.052\,s. For the peak around 272\,s, we found that 
the 271.95\,s peak matched our light curves better than the one at 270.455\,s 
(see Fig. \ref{fig:F2cfperiods} and the lowermost curve in each of the panels
of Fig. \ref{fig:glcft}). Therefore, even though \citet{kepler:95} assigned a 
lower significance to the former than to the latter we chose to fix the
period of this peak to 271.95\,s. We have checked that this choice is of
no relevance to what follows. The amplitudes and phases of the modulations in 
each light curve are listed in Table \ref{gtab}. 

Fig. \ref{fig:glcft} shows that the periodicity at 215\,s is clearly detected at 
all wavelengths, while those at 272 and 304\,s are more difficult to discern in the 
ultra-violet datasets (Fig. \ref{fig:glcft}). The phases for each of the modes are 
identical within the errors for the optical dataset. This is to be expected if the 
variations in brightness are dominated by variations in temperature \citep{rkn:82}. 
This is not the case for the ultra-violet datasets presented here, as the data were 
not acquired simultaneously in the different passbands i.e. the phases in Table 
\ref{gtab} are given with respect to the (different) zero points for each run.

From Fig. \ref{fig:glcft} and Table \ref{gtab}, it is clear that the
behaviour of F2 is markedly different from that of either F1 and F3 in
that its amplitude increases by more than expected if it were to be
adequately described by a low spherical degree mode.

\begin{figure}[t]
\plotone{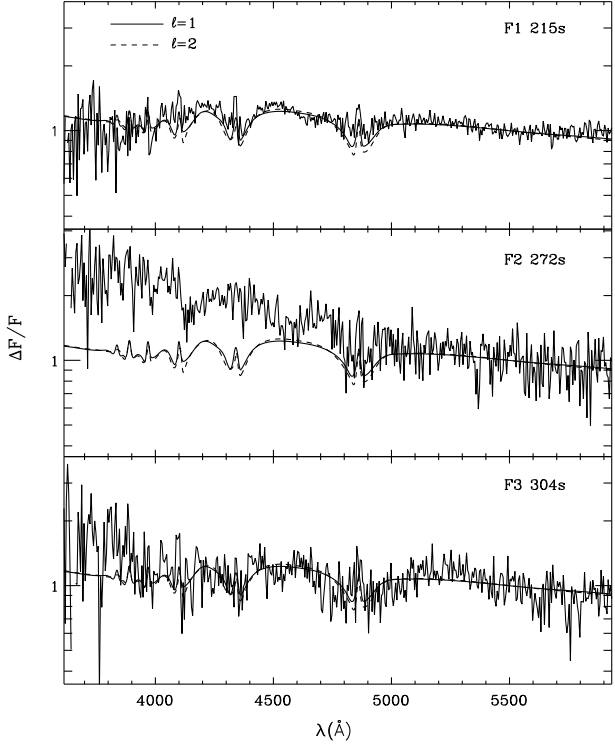}
\caption{Fractional wavelength dependent pulsation amplitudes of each of the
3 modes calculated in 5\,{\AA} wide bins. The period is indicated next to the
name. The observations have been overlaid with model chromatic amplitudes with
\Teff\ =11.5\,kK and $\log g=8$ with ML2/$\alpha$=0.6. Both the observations
and the models have been normalised to unity at 5500\,{\AA}. The models have
been convolved with a Gaussian function (FWHM=4.4\,{\AA}) to emulate a seeing
profile. See Sect. \ref{subsec:champF1} for details of the parameters used in 
computing the models.}
\label{fig:champs}
\end{figure}

In order to check that any dependence on wavelength was not introduced 
by the way in which the light curves were constructed, we deselected the
$-$3500\,$<$\,t(s)$<$\,$-$2500 region of each of the lightcurves (see Fig. 
\ref{fig:goptlcs}) -- which shows somewhat similar modulations for the 
4585 and 4210\,{\AA} light curves -- and fit these light curves in the 
manner described above. We found that the amplitude of F2 only changed by 
0.02, 0.04, and 0.06\%  going from the longest to shortest wavelength light 
curves. These changes are insignificant compared to the errors attached to 
the measurements which are nearly identical to those listed in Table \ref{gtab}.

In principle, a varying background could give rise to an additional 
wavelength-dependence in the light curves that may have the effect of
artificially enhancing mode amplitudes. To test this possibility, we
constructed light curves of the background in each of the three passbands.
We do not find any peaks in the Fourier Transforms of these light curves
that are coincident with the modulations we find in G 117-B15A.
We test this more rigorously in Sect. \ref{subsec:weird272}.

\section{Chromatic amplitudes and phases}

\begin{figure}[t]
\plotone{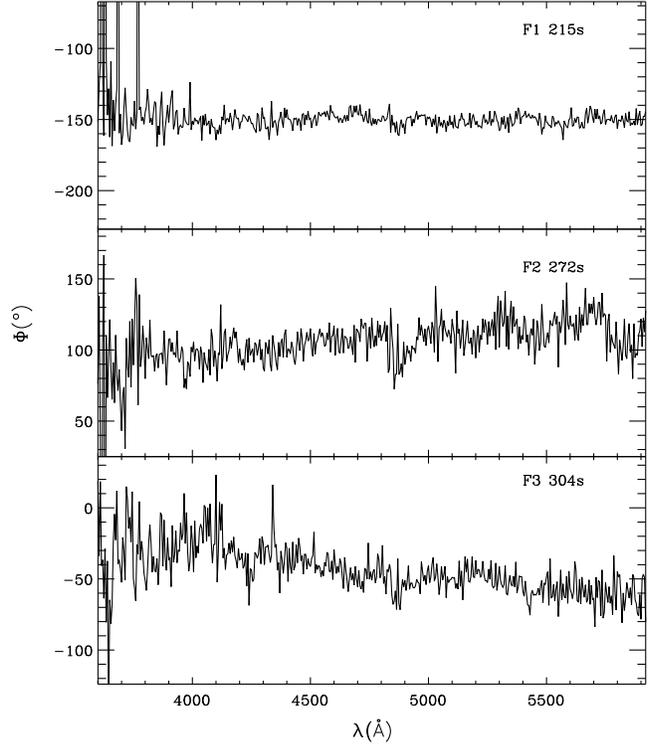}
\caption{Chromatic phases. The slopes in the continuum phases of F2 
and F3 are unexpected. Based on simulations, however, we believe
that this might not be intrinsic to the star. This, in contrast to the
unexpected deviations from the models of the chromatic amplitudes for
F2 (Fig. \ref{fig:champs}; see text).}
\label{fig:chrph}
\end{figure}

As mentioned earlier, the fractional wavelength-dependent pulsation 
amplitudes (i.e. ``chromatic amplitudes'') can be used to infer the 
$\ell$ value of a mode. The technique is based on the expectation 
that although all modes suffer cancellation due to integration over the 
stellar disc, this cancellation is diminished at shorter wavelengths due 
to limb-darkening, and is a function of the $\ell$ character of a mode. 
Thus, not only are the pulsation amplitudes larger at shorter wavelengths, 
but higher $\ell$ modes ($\ell \la 3$) can potentially be observed. 
At optical wavelengths, smaller effects within the Balmer lines are also 
apparent.

Chromatic amplitudes for F1, F2, and F3 are shown in Fig. \ref{fig:champs}. 
They were calculated by fitting the three periodicities in each 5\,{\AA}
bin, and by fixing the frequencies to those listed in Table \ref{gtab},
but leaving the amplitudes and phases free to vary. The resulting chromatic
phases are shown in Fig. \ref{fig:chrph}. Fixing the phases to the values 
in Table \ref{gtab} makes almost no difference to the shape of the chromatic 
amplitudes, not even for F3, which shows a fairly large variation in phase.

Synthetic chromatic amplitudes, computed using model atmospheres kindly provided 
by D. Koester, and having $\ell=1$ and $\ell=2$ are overlaid on Fig. \ref{fig:champs}
for \Teff$\thickspace=11.5$\,kK. Compared to these models, as well as chromatic 
amplitudes of other ZZ Cetis, that of F1 closely matches the models, while that 
of F2 shows a pronounced slope and a lack of distinct line cores; the chromatic 
amplitude of F3 is noisier and consistent with both $\ell=1$ and $\ell=2$ models.
That F2 is rather different from F1 and F3 can also be appreciated from 
the Fourier transforms shown in Fig. \ref{fig:glcft}, where it is much stronger 
at 4210\,\AA, relative to its strength at 5500\,\AA, than either F1 or F3.
We discuss the constraints on spherical degree and effective temperature that
can be placed using the chromatic amplitude of F1 in Sect. \ref{subsec:champF1}
below.

We now turn to the chromatic phases (Fig. \ref{fig:chrph}). As mentioned
earlier, the continuum phases are expected to be constant with wavelength.
Although this is the case for F1, both F2 and F3 show large deviations --
of the order of tens of degrees compared to 1-2\arcdeg\ found in previous
studies \citep[e.g. \object{ZZ Psc},][]{cvkw:00}. We argue below (Sect. 
\ref{subsec:weird272}) that these variations are most likely due to noise.

\subsection{Spherical degree of F1 and the effective temperature of G 117-B15A}
\label{subsec:champF1}

\begin{figure}[!t]
\plotone{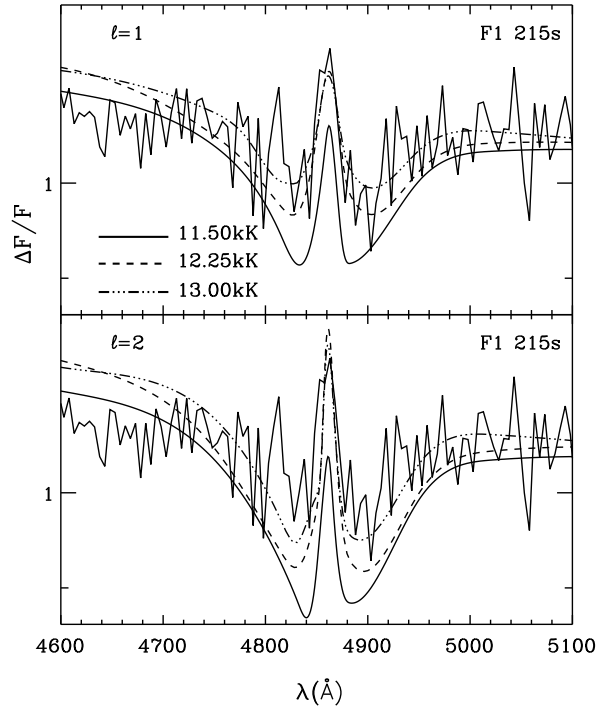}
\caption{Expanded view of the chromatic amplitudes and models 
around H$\beta$. The top panel shows the strongest mode, F1, overlaid 
with models having $\ell=1$ and three different effective temperatures, 
as indicated. The bottom panel shows the same models for $\ell=2$. 
The 13.0\,kK model fails to match the observations for both $\ell$ 
values. The $\ell=1$ models are a better match at either of the
other two effective temperatures considered than the $\ell=2$ 
models. This is apparent both in the wings and in the line cores. 
Note however, that for our grid of models, the 12.25\,kK, $\ell=1$ 
model is a better match in the wings of the lines than the corresponding 
11.5\,kK model.}
\label{fig:zoomhb}
\end{figure}

Given the faintness of G 117-B15A (V=15.54) relative to other objects which 
have been studied using this technique \citep[e.g. ZZ Psc, V=13.0;][]{cvkw:00},
and the correspondingly lower signal-to-noise ratio, it is difficult to 
distinguish between $\ell=1$ and $\ell=2$ using the models in the continuum 
regions, even for the strongest mode (F1). However, the $\ell=1$ model is a 
better match in the wings of the line cores than the $\ell=2$ model (see 
Fig. \ref{fig:zoomhb}). This is true even for models with different effective 
temperatures.

We can use the chromatic amplitude of F1 to constrain the effective
temperature. This is useful, since model spectra of pulsating DA
white dwarfs are not very sensitive to small changes in the atmospheric 
parameters, and fits to the average spectrum almost never yield unique
results. This is because the Balmer lines reach their maximum strength 
in the ZZ Ceti instability strip. Other constraints, e.g. ultra-violet 
spectra or parallaxes are necessary to pin down the atmospheric 
parameters \citep{kv:96}. 

\citet{berg:95} find that the optical and ultra-violet observations are
best reproduced with the ML2 mixing length prescription, with the mixing
length parameter $\alpha=0.6$. For G 117-B15A, they infer
\Teff$\thickspace=11620$\,K and $\log g=7.97$. The models used here have 
also been computed using the above prescription. From fits (not shown)
to the Balmer lines in our average spectrum, we infer $\sim\!11750\,$K,
consistent with \citet{berg:95}.
Our chromatic amplitudes of F1, however, can be reproduced 
better using a higher temperature, of $\sim\!122250\,$K. A similar 
difference was found for ZZ Psc by \citet{cvkw:00} and for the DBV, 
\object{GD 358} by \citet{kotak:03}. This may indicate flaws in the model 
atmospheres or in the way we use them to calculate chromatic amplitudes. 
Below, we adopt $T_{\rm eff}=11620\,$K where necessary, but stress that 
this choice has little bearing on what follows.

\subsection{Is the 272\,s mode intrinsically weird?}
\label{subsec:weird272}

\begin{figure}
\plotone{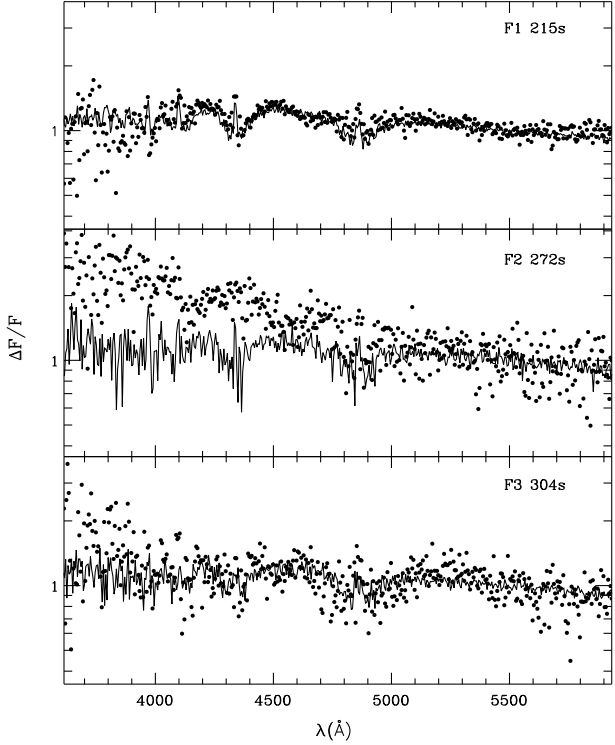}
\caption{Observed chromatic amplitudes (dots) and synthetic chromatic amplitudes
(full line) with noise added and treated in exactly the same way as the data 
(see Sect. \ref{sec:periodlc} for details).} 
\label{fig:simchamps}
\end{figure}

The chromatic amplitude of F2 is strikingly different from that expected.
We carried out two tests in order to to assess the contribution due to noise
to the chromatic amplitude of F2. We first generated synthetic chromatic 
amplitudes using model spectra and the observed light curve. This was done as 
follows: first, for every point in the 5500\,{\AA} light curve, we determined 
a temperature by simple scaling, taking the mean temperature to be 11.62\,kK. 
Next, we interpolated in a grid of model spectra (all with $\log g=8$) to find 
a simulated spectrum for this point, and added random, normally-distributed noise 
(adding more noise at shorter wavelengths). Finally, these simulated spectra were
used to calculate chromatic amplitudes. An example is shown in Fig. \ref{fig:simchamps}
overlaid on the observations. Two points are evident: first, it is conceivable
that the line cores can be diminished or enhanced by random noise peaks. Secondly,
normally-distributed measurement noise alone cannot reproduce the slope of F2.

The slope shown by the chromatic phases of F2 and F3 is not reproduced 
(in either direction) by our simulations and cannot therefore be due to noise
alone. 

\begin{table}[!t]
\caption[]{Result of Monte Carlo simulation}
\label{tab:mcfake}
\begin{center}
\begin{tabular}{cccc}
\hline \noalign{\smallskip}
   Central $\lambda$      & fake\,1      & fake\,2  &       \\
      ({\AA})             & (\%)         & (\%)     &       \\ \noalign{\smallskip}
\hline \noalign{\smallskip}
      5500                &  51.5        &  53.4     &       \\
      4585                &  1.8         &  1.2      &        \\
      4210                &  $<$1        &  $<$1     &        \\   \noalign{\smallskip}
\hline
\end{tabular}
\end{center}
\tablecomments{fake\,1 and fake\,2 refer to the fake signals inserted between 
 2000-3000\,$\mu$Hz and 5500-6500\,$\mu$Hz respectively. The second and 
 third columns show the fraction of trials (out of 1000) that either artificial
 peak had an amplitude at least as large as that measured in the corresponding
 passband for F2.}
\end{table}

\begin{figure}[!t]
\plotone{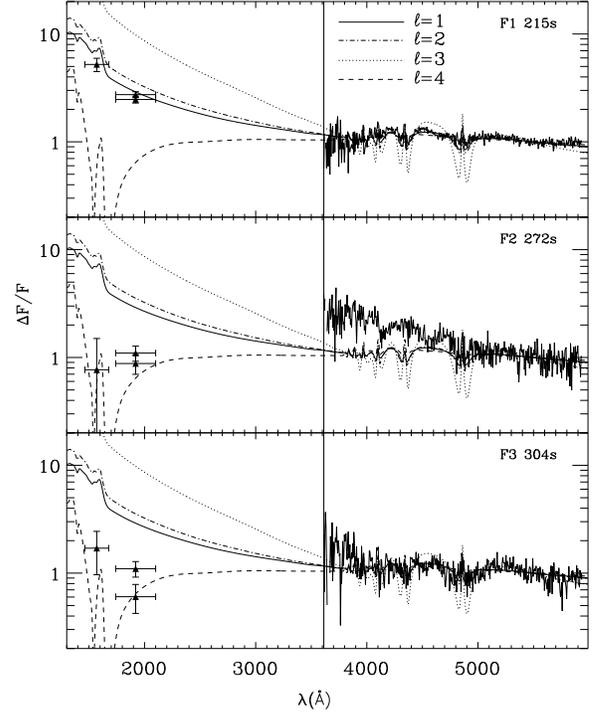}
\caption{Observed and model chromatic amplitudes for the whole wavelength 
range. The model shown here is the same as in Fig. \ref{fig:champs} i.e. 
\Teff$\thickspace=11.5$\,kK, $\log g=8$, and ML2/$\alpha=0.6$.
The amplitudes have been normalised to unity at 5500\,{\AA}. The
horizontal error bars on the ultra-violet amplitudes represent the FWHM 
of the filters that were used. The vertical line merely serves to emphasise 
that the optical and ultra-violet observations are not contemporaneous.}
\label{fig:uvoptchamp_zz11500}
\end{figure}

The above test rests on the assumption that the noise in our data is normally
distributed. This may not be the case. For instance, instrumental effects could
lead to variations that are coherent with wavelength. Most effects we can think 
of would be multiplicative and grey, and would seem unlikely to cause the deviant
chromatic amplitudes seen for F2. In principle, however, it may be possible to
conceive of a noise signal, perhaps an additive one, that is much stronger in the 
blue. To test whether such signals are present, and whether they could lead to 
chromatic amplitudes such as those observed for F2, we performed another test,
now using each of the three optical passbands as follows. We first injected two 
sinusoidal signals in the 5500\,{\AA} passband with arbitrary phases (between 0 
and 2$\pi$ radians) and frequencies randomly chosen to lie between 2000-3000\,$\mu$Hz 
and between 5500-6500\,$\mu$Hz which are regions free of any obvious real signal 
(see top panel of Fig. \ref{fig:glcft}). Both amplitudes were chosen to be equal to 
that of F2 as listed in Table \ref{gtab}.
Next, we inserted these two signals in the two other passbands (4585\,{\AA}, 
4210\,{\AA}), with the same phase and frequency, but with the amplitudes scaled to the
ratios measured for F1 i.e. F1$_{4585}$/F1$_{5500}=1.29$ and F1$_{4210}$/F1$_{5500}=
1.36$. This allows us to establish whether or not random noise peaks can artificially 
enhance the amplitude at shorter wavelengths in a manner similar to that seen for F2. 
We repeated this procedure 1000 times per passband and measured the amplitudes of 
the fake signals in exactly the same way as described in Sect. \ref{sec:periodlc} for 
the real data.  If the anomalous amplitudes of F2 shortwards of $\sim$\,5000\,{\AA}
are due to noise, then we expect to find that the artificial peaks have an amplitude
at least as large as that of F2 in a large percentage of the trials. The outcome of 
this test is shown in Table \ref{tab:mcfake}. The results for the 4585\,{\AA} and 
4210\,{\AA} regions confirm that the measured amplitude of F2 is larger than can be
expected from the contribution of noise alone. (For the 5500\,{\AA} passband, the
values near 50\% merely reflect the fact that we inserted signals at the strength
of F2 and that noise will almost as often decrease as increase this signal).

The absolute difference in phases between the 5500\,{\AA} and 4210\,{\AA} passbands is 
as large as that observed for the chromatic phase of F3 ($\sim\negthickspace\,25\arcdeg$)
13.6\% and 7.6\% of the trials for the peaks labelled fake1 and fake2 respectively.
These values show that there is certainly some contribution due to noise to the slope 
of the chromatic phases of F2 and F3. The different contributions indicated by fake1 
and fake2 probably reflect the fact that the noise level at the lower frequency end (fake1) 
shows a somewhat greater increase from 5500\,{\AA}--4210\,{\AA} than at the higher frequency 
end (fake2); this is seen in Fig. \ref{fig:glcft}.

We conclude that the change in phase with wavelength shown by F2 may well be due 
to noise, but that the strange chromatic amplitudes are most likely intrinsic.
F3 is simply too noisy to make a meaningful comparison.

\subsection{Constraints from the ultra-violet light curves}

Fig. \ref{fig:uvoptchamp_zz11500} shows the constraints obtained from ultra-violet 
photometry. Again, while F1 behaves as expected, i.e. larger amplitudes at 
shorter wavelengths, both F2 and F3 have smaller amplitudes in the ultra-violet than 
in the optical regions. For modes having $\ell \le 2$, i.e. modes which are 
most likely to be observable, this behaviour is entirely unexpected. Model 
chromatic amplitudes for \Teff$\thickspace=11.5$\,kK show that only an $\ell=4$ 
model can match the data for F2 and F3. 

As the optical and ultraviolet observations were not taken
simultaneously, one might appeal to mode variability to explain the
unexpected mismatch between optical and ultraviolet.  For F3, we
cannot exclude this possibility. For F2, however, the relative
amplitudes in the ultraviolet, which were determined
quasi-simultaneously, are also inconsistent with the predicted ones.

\section{Discussion}

In spite of the anomalous behaviour of F2, we continue, for the time 
being, to assume that this modulation seen in both the optical and 
ultra-violet light curves is due to pulsation. We base this assumption on 
the fact that the periods of F2 (272\,s) and F3 (304\,s) lie comfortably in the 
regime of non-radial $g$-mode oscillations, and that similar periods have been 
observed in several other ZZ Cetis. For instance, ZZ Ceti itself shows three 
dominant periods at 215, 271, and 304\,s \citep{kepler:82}.

We have only considered temperature variations when computing our synthetic
chromatic amplitudes. It is well-known that these completely dominate all 
other sources of luminosity variations \citep{rkn:82}. Any other variation 
would be expected to be independent of wavelength to first order.
Also, any peculiar limb darkening law is at odds with those obtained from 
model atmospheres. 

We briefly entertain the possibility that the that the 272\,s modulation 
is actually a combination mode generated by real modes that are rendered
invisible. This happens, for example, for the first harmonic of an 
$(\ell,m)=(1,0)$ mode as the inclination approaches 90\arcdeg, 
and for the first harmonic of $(\ell,m)=(2,0)$ mode at intermediate 
(40-65\arcdeg) inclinations (see Fig. 2 in Wu, 2001 for $\ell=m$ modes
or Fig. 4.9 in Kotak 2002 for $\ell\ne m$ modes).
However, the flux variations are exceedingly low due to the cancellation 
suffered by high $\ell$ modes, even for a fortuitous inclination
angle. As the slope of the chromatic amplitude in the optical cannot 
be explained by the above argument, this conjecture too must be discarded.

An intriguing possibility is that the 272\,s mode is not a $g$-mode at 
all. The studies of \citet{saio:82} and \citet{berthprov:83} showed that 
$r$-modes could be excited in ZZ Ceti type variables with periods similar 
to those of $g$-modes. However, to first order in $\Omega$ (the angular 
frequency of rotation), an $r$-mode produces no brightness changes 
\citep{saio:82,kepler:84}. An estimate of $v\sin i$ for G 117-B15A 
is not available, but other white dwarfs have been found to rotate 
with periods of about one day. Unless G 117-B15A is an exceptionally 
fast rotator, $r$-modes are probably do not give rise to F2. Estimates 
of the contamination of $g$-modes by $r$-modes cannot be established 
without detailed modelling, but it is difficult to see why these should 
be wavelength-dependent.

The nature of F2 unfortunately remains unexplained as all global 
effects must affect all modes. Either strong nonlinear processes 
may have to be invoked, or the description of modes based on a single 
spherical harmonic may have to be called into question. 

We end by pointing out that the behaviour of F2 and F3 in the ultra-violet 
is not unique to G 117-B15A; the 141\,s mode in \object{G 185-32} also shows 
similar behaviour \citep{kepler:00b}. One other similarity between F2 and F3 
and the 141\,s mode of G 185-32 is that they are the lowest amplitude modulations 
present in the spectrum. Simultaneous ultra-violet and optical spectroscopy 
might help to clarify the nature of the puzzling variations in G 117-B15A 
and G 185-32. Since both G 117-B15A and G 185-32 are relatively close to the 
blue edge, it might be relevant to establish whether such behaviour is peculiar 
to white dwarfs just entering the ZZ Ceti instability strip.
Our findings highlight the need for caution and mode identification by several 
different methods. Indeed, given the above, any attempts to measure the rate of
change of period or to constrain the core composition based on either the 272\,s 
or 304\,s modes may prove to be futile.

\begin{acknowledgements}
R.K. would like to sincerely thank H-G. Ludwig for his interest 
and many fruitful discussions. R.K. would also like to thank the
Royal Commission for the Exhibition of 1851 for partial support.
Based in part on observations made with the NASA/ESA Hubble Space
Telescope, obtained from the data archive at the Space Telescope Science
Institute. STScI is operated by the Association of Universities for
Research in Astronomy, Inc. under NASA contract NAS 5-26555.
\end{acknowledgements}


\begin{thebibliography}{}

\bibitem[\protect\citeauthoryear{Bergeron et al.}{1995}]{berg:95}
 Bergeron P., Wesemael F., et al., 1995 ApJ, 449, 258

\bibitem[\protect\citeauthoryear{Berthomieu \& Provost}{1983}]{berthprov:83}
 Berthomieu, G., \& Provost, J., 1983, A\&A, 122, 199

\bibitem[\protect\citeauthoryear{Bradley}{1998}]{bradley:98}
 Bradley, P.A., 1998, ApJSS, 116, 307

\bibitem[\protect\citeauthoryear{Clemens et al.}{2000}]{cvkw:00}
 Clemens, J. C., van Kerkwijk, M. H., \& Wu, Y., 2000, MNRAS, 314, 220

\bibitem[\protect\citeauthoryear{Finley et al.}{1997}]{fin:97}
Finley, D.S., Koester D., \& Basri G., 1997, ApJ, 488, 375

\bibitem[\protect\citeauthoryear{Hauschildt et al.}{1999}]{hauschildt:99}
Hauschildt, P.H., Allard, F., Ferguson, J., et al. 1999, ApJ 525, 871

\bibitem[\protect\citeauthoryear{Kepler}{1984}]{kepler:84}
 Kepler, S.O., 1984, ApJ, 286, 314

\bibitem[\protect\citeauthoryear{Kepler et al.}{2000a}]{kepler:00a}
 Kepler, S.O., Mukadam A., Winget D.E. et al., 2000a, ApJ, 534, L185

\bibitem[\protect\citeauthoryear{Kepler et al.}{2000b}]{kepler:00b}
 Kepler, S.O., Robinson, E.L., Koester, D. et al., 2000b, ApJ, 539, 379

\bibitem[\protect\citeauthoryear{Kepler et al.}{1982}]{kepler:82}
 Kepler, S.O., Robinson, E.L., Nather, R. E., McGraw, J. T. 1982, ApJ, 254, 676

\bibitem[\protect\citeauthoryear{Kepler et al.}{1995}]{kepler:95}
 Kepler, S.O., Winget D.E., Nather, R.E. et al., 1995, Balt. Astr., 4, 221

\bibitem[\protect\citeauthoryear{Koester}{2002}]{koester:02}
 Koester, D., 2002, A\&ARv, 11, 33

\bibitem[\protect\citeauthoryear{Koester et al.}{1994}]{koester:94}
 Koester, D., Allard, N.F., \& Vauclair, G., 1994, A\&A, 291, L9

\bibitem[\protect\citeauthoryear{Koester \& Vauclair}{1996}]{kv:96}
  Koester D., \& Vauclair G. 1996, White Dwarfs, eds. J. Isern, M Hernanz,
  E. Garcia-Berro (Kluwer), 429

\bibitem[\protect\citeauthoryear{Kotak}{2002}]{kotakthesis:02}
 Kotak, R., 2002, Thesis, Lund Observatory, Sweden

\bibitem[\protect\citeauthoryear{Kotak et al.}{2002}]{kotakhs:02}
 Kotak, R., van Kerkwijk, M.H., \& Clemens, J.C., 2002 A\&A, 388, 219

\bibitem[\protect\citeauthoryear{Kotak et al.}{2003}]{kotak:03}
 Kotak, R., van Kerkwijk, M. H., Clemens, J. C., Koester, D, 
 2003, A\&A, 397, 1043

\bibitem[\protect\citeauthoryear{McGraw \& Robinson}{1976}]{mcgrob:76}
 McGraw, J.T., \& Robinson, E.L., 1976, ApJ, 200, L89

\bibitem[\protect\citeauthoryear{Nather et al.}{1990}]{nath:90}
  Nather, R.E., Winget, D.E., Clemens, J.C. et al., 1990, ApJ, 361, 309

\bibitem[\protect\citeauthoryear{Oke et al.}{1995}]{oke:95}
  Oke, J.B., Cohen, J.G., Carr, et al. 1995, PASP, 107, 375

\bibitem[\protect\citeauthoryear{Richer \& Ulrych}{1974}]{richul:74}
 Richer, H.B., \& Ulrych, T. J., 1974, ApJ, 192, 719

\bibitem[\protect\citeauthoryear{Robinson et al.}{1982}]{rkn:82}
  Robinson E., Kepler S., Nather E., 1982, ApJ, 259, 219

\bibitem[\protect\citeauthoryear{Robinson et al.}{1995}]{robetc:95}
  Robinson E., Mailloux T., Zhang E., et al.  1995, ApJ, 438, 908

\bibitem[\protect\citeauthoryear{Saio}{1982}]{saio:82}
 Saio, H., 1982, ApJ, 256, 717

\bibitem[\protect\citeauthoryear{Wu}{2001}]{wu:01}
 Wu, Y., 2001, MNRAS, 323, 248

\end{thebibliography}
\end{document}